\documentclass[aps,prc,preprint,groupedaddress,showpacs,superscriptaddress,floatfix]{revtex4-1}

\usepackage{amsmath}\allowdisplaybreaks[1]
\usepackage{graphicx}
\usepackage{amssymb}
\usepackage{bm}
\usepackage{array}
\usepackage[colorlinks,linkcolor=blue,urlcolor=blue,anchorcolor=blue,citecolor=blue]{hyperref}
\usepackage{slashed}
\usepackage{braket}
\usepackage{xcolor}
\usepackage{tabularx}
\newcommand*{\DOI}[2]{\href{http://dx.doi.org/#1}{#2}}

\begin{document}

\title{Search for a hidden strange baryon-meson bound state from \texorpdfstring{$\phi$}{phi} production in a nuclear medium}

\newcommand*{\DUKE}{Department of Physics, Duke University and Triangle Universities Nuclear Laboratory, Durham, North Carolina 27708, USA}\affiliation{\DUKE}
\newcommand*{\DKU}{Duke Kunshan University, Kunshan, Jiangsu 215316, China}\affiliation{\DKU}
\newcommand*{\NJNU}{Department of Physics, Nanjing Normal University, Nanjing, Jiangsu 210097, China}\affiliation{\NJNU}
\newcommand*{\NJU}{Department of Physics, Nanjing University, Nanjing, Jiangsu 210093, China}\affiliation{\NJU}

\author{Haiyan Gao}\affiliation{\DUKE}\affiliation{\DKU}
\author{Hongxia Huang}\email{hxhuang@njnu.edu.cn}\affiliation{\DUKE}\affiliation{\NJNU}
\author{Tianbo Liu}\email{liutb@jlab.org}\affiliation{\DUKE}\affiliation{\DKU}
\author{Jialun Ping}\affiliation{\NJNU}
\author{Fan Wang}\affiliation{\NJU}
\author{Zhiwen Zhao}\affiliation{\DUKE}


\begin{abstract}
  We investigate the hidden strange light baryon-meson system. With the resonating-group method, two bound states, $\eta'-N$ and $\phi-N$, are found in the quark delocalization color screening model. Focusing on the $\phi-N$ bound state around 1950\,MeV, we obtain the total decay width of about 4\,MeV by calculating the phase shifts in the resonance scattering processes. To study the feasibility of an experimental search for the $\phi-N$ bound state, we perform a Monte Carlo simulation of the bound state production with an electron beam and a gold target. In the simulation, we use the CLAS12 detector with the Forward Tagger and the BONUS12 detector in Hall B at Jefferson Lab. Both the signal and the background channels are estimated. We demonstrate that the signal events can be separated from the background with some momentum cuts. Therefore it is feasible to experimentally search for the $\phi-N$ bound state through the near threshold $\phi$ meson production from heavy nuclei.
\end{abstract}


\maketitle

\section{Introduction}

The study of multiquark states is one of the most active frontiers since the establishment of the quark model by Gell-Mann and Zweig~\cite{GellMann:1964nj,Zweig:1981pd}. Jaffe was the first to carry out quantitative studies~\cite{Jaffe:1976ig}, and Lipkin extended the idea to pentaquarks~\cite{Lipkin:1987sk}. As the fundamental theory of the strong interaction, the quantum chromodynamics (QCD) does not forbid the existence of pentaquark states. Recently, the discovery of hidden charm pentaquark candidates was reported by LHCb~\cite{Aaij:2015tga}, and it invoked a renewed interest in this field~\cite{Chen:2016qju}. 

It is pointed out by Brodsky, Schmidt, and de~T\'{e}ramond that the QCD van de Waals interaction, mediated by multigluon exchanges, will dominate the interaction between two hadrons when they have no common quarks~\cite{Brodsky:1989jd}. It was further shown by Luke, Manohar, and Savage that the QCD van de Waals force is enhanced at low relative velocities between the two~\cite{Luke:1992tm}. This finding supports the prediction that a nucleon/nucleus-charmonium bound state can be produced near the charm production threshold. As an analogy, a $\phi-N$ bound state is predicted by Gao, Lee, and Marinov~\cite{Gao:2000az}. It is also pointed out that the subthreshold quasifree $\phi$ meson photoproduction inside a nuclear medium will enhance the probability for the formation of the $\phi-N$ bound state. In addition, some chiral quark model calculation~\cite{Huang:2005gw} and lattice QCD calculation~\cite{Beane:2014sda} in recent years also support the existence of such a kind of bound state. On the other hand, the $\phi-N$ bound state can be viewed as a hidden strange pentaquark state. A comparison with the hidden charm pentaquark candidates will unveil the flavor-dependent effect in hadron physics. Thus the experimental search for the $\phi-N$ bound state is of great interest and will improve our understanding of the strong interaction.

In this paper, we carry out a front-to-end study of the search for the $\phi-N$ bound state. We perform a calculation with the quark delocalization color screening model (QDCSM)~\cite{Wang:1992wi,Wu:1996fm,Pang:2001xx}, which is developed aiming to understand the similarities between nuclear and molecular forces despite different scales. In this model, the intermediate-range attraction is achieved by the quark delocalization, which is like the electron percolation in molecules. The color screening provides an effective description of the hidden color channel coupling~\cite{Huang:2011kf}, and leads to the possibility of the quark delocalization. The QDCSM was utilized to investigate the baryon-baryon scattering phase shifts in the framework of the resonating group method (RGM). It provides a good description of the nucleon-nucleon and nucleon-hyperon interactions and the deuteron properties~\cite{Ping:2000dx,Ping:1998si,Wu:1998wu,Pang:2001xx}. Some dibaryon candidates are also studied with this model~\cite{Ping:2008tp,Chen:2011zzb}. The one of particular interest is a narrow resonance $N-\Omega$ state~\cite{Huang:2015yza}, which is proposed for searches in heavy ion collisions and a hadron beam experiment with a newly developed automatic scanning system~\cite{Yoshida2015}. Moreover the hidden charm pendaquark candidates, $\eta_c-N$ and $J/\psi-N$ bound states, are also studied in QDCSM~\cite{Huang:2015uda}. As the strangeness counterpart, we study the hidden strange system with a light baryon and a light meson in the QDCSM. Some baryon-meson bound states are found from the model calculation, and one of them is the $\phi-N$ bound state. The decay properties of this bound state are calculated from the phase shifts of the resonance scatterings. To investigate the feasibility of an experimental search for this bound state, we study the production of this bound state from the near threshold $\phi$ production process in a nuclear medium as suggested in~\cite{Gao:2000az}. Particularly, we choose a gold target as an example, and perform a Monte Carlo simulation with electron beams at Jefferson Lab and detectors in Hall B. Events with scattered electrons detected by the Forward Tagger~\cite{ForwardTagger} and $pK^+K^-$ detected by the CLAS12~\cite{CLAS12} and the BONUS12~\cite{BONUS12} are selected to reconstruct the bound state. Both the signal and the background are estimated in the simulation. With a set of momentum cuts motivated by the previous study~\cite{Liska:2007de}, we demonstrate that the signal events can be separated from the background. Therefore it is possible to search for the $\phi-N$ bound state in experiments.

The paper is organized as follows. In Sec.~\ref{sec2}, we briefly introduce the quark delocalization color screening model, and then we calculate the properties of the hidden strange baryon-meson system in Sec.~\ref{sec3}. Focusing on the $\phi-N$ bound state, we investigate the production process, and perform a simulation to show the feasibility of the experimental search for this state in Sec.~\ref{sec4}. The discussion and the conclusion are drawn in the last section.

\section{Quark delocalization color screening model \label{sec2}}

As described in~\cite{Wang:1992wi,Wu:1996fm,Pang:2001xx,Ping:1998si,Wu:1998wu,Pang:2001xx,Ping:2008tp,Chen:2011zzb}, the Hamiltonian of QDCSM is expressed as
\begin{equation}
H = \sum_{i=1}^5 \left(m_i+\frac{\bm{p}_i^2}{2m_i}\right) -T_c
+\sum_{i<j} \left[ V^{G}(r_{ij})+V^{\chi}(r_{ij})+V^{C}(r_{ij})
\right],
\end{equation}
where
\begin{align}
V^{G}(r_{ij})&= \frac{1}{4}\alpha_{s} \bm{\lambda}_i \cdot
\bm{\lambda}_j
\left[\frac{1}{r_{ij}}-\frac{\pi}{2}\left(\frac{1}{m_{i}^{2}}+\frac{1}{m_{j}^{2}}+\frac{4\bm{\sigma}_i\cdot\bm{\sigma}_j}{3m_{i}m_{j}}\right)
\delta(r_{ij})-\frac{3}{4m_{i}m_{j}r^3_{ij}}S_{ij}\right], \\
V^{\chi}(r_{ij})&= \frac{1}{3}\alpha_{\rm ch}
\frac{\Lambda_{\chi}^2}{\Lambda_{\chi}^2-m_{\chi}^2}m_\chi \left\{ \left[
Y(m_\chi r_{ij})- \frac{\Lambda_\chi^3}{m_{\chi}^3}Y(\Lambda_\chi r_{ij})
\right]\bm{\sigma}_i\cdot\bm{\sigma}_j \right.\nonumber \\
&\quad \left. +\left[ H(m_\chi r_{ij})-\frac{\Lambda_\chi^3}{m_\chi^3}
H(\Lambda_\chi r_{ij})\right] S_{ij} \right\} \bm{\tau}_i\cdot\bm{\tau}_j, ~~~\chi=\pi,K,\eta, \\
V^{C}(r_{ij})&= -a_c \bm{\lambda}_i\cdot\bm{\lambda}_j [f(r_{ij})+V_0], \\
 f(r_{ij}) & = \left\{ \begin{array}{ll}
 r_{ij}^2 &
 \qquad \mbox{if }i,j\mbox{ occur in the same baryon orbit} \\
  \frac{1 - e^{-\mu_{ij} r_{ij}^2} }{\mu_{ij}} & \qquad
 \mbox{if }i,j\mbox{ occur in different baryon orbits} \\
 \end{array} \right.
\nonumber
\end{align}
The $S_{ij}$ is the quark tensor operator:
\begin{equation}
S_{ij} = \frac{(\bm{\sigma}_i\cdot\bm{r}_{ij})
(\bm{\sigma}_j\cdot \bm{r}_{ij})}{r_{ij}^2}-\frac{1}{3}\bm{\sigma}_i \cdot\bm{\sigma}_j,
\end{equation}
and the subscripts $i,j$ denote the quark index in the system. The $\bm{\sigma}$ and the $\bm{\tau}$ are Pauli matrices that, respectively, describe the spin and the isospin spaces, and the $\bm{\lambda}$s are the Gell-Mann matrices that describe the color degrees of freedom. The $Y(x)$ and $H(x)$ are the standard Yukawa functions~\cite{Valcarce:2005em}, the $T_c$ is the center-of-mass kinetic energy, the $\alpha_{\rm ch}$ is the chiral coupling constant which is usually determined from the $\pi N$ scatterings, the $\Lambda_\chi$ is the chiral symmetry breaking scale, and the $\alpha_{s}$ is the quark-gluon strong coupling constant. To cover the energy range from up and down quarks to strange quarks, one can introduce an effective running coupling as~\cite{Vijande:2004he}
\begin{equation}
\alpha_{s}(\mu) = \frac{\alpha_{s}(\mu_0)}{\ln\frac{\mu^2+\mu_0^2}{\Lambda_{0}^2}}.
\end{equation}
In the phenomenological confinement potential $V^C$, the color screening parameter $\mu_{ij}$ is determined by fitting the deuteron properties, $NN$ scattering phase shifts, and $N\Lambda$ and $N\Sigma$ scattering cross sections as $\mu_{qq}=0.45\,\mathrm{fm}^{-2}$ and $\mu_{ss}=0.08\,\mathrm{fm}^{-2}$ where $q$ represents $u$ or $d$ and $\mu_{qs}$ is constrained by $\mu_{qs}^2=\mu_{qq}\mu_{ss}$. 

The quark delocalization effect is realized by specifying the single-particle orbital wave function in QDCSM as a linear combination of left and right Gaussians as
\begin{align}
\psi_{\alpha}(\bm{s}_i ,\epsilon) & = \left(\phi_{\alpha}(\bm{s}_i)
+ \epsilon \phi_{\alpha}(-\bm{s}_i)\right) /N(\epsilon), \\
\psi_{\beta}(-\bm{s}_i ,\epsilon) & =\left(\phi_{\beta}(-\bm{s}_i)
+ \epsilon \phi_{\beta}(\bm{s}_i)\right) /N(\epsilon),
\end{align}
where
\begin{align}
N(\epsilon) & = \sqrt{1+\epsilon^2+2\epsilon e^{-s_i^2/4b^2}}, \label{1q} \\
\phi_{\alpha}(\bm{s}_i) & = \left( \frac{1}{\pi b^2}\right)^{3/4} e^{-\frac{1}{2b^2} (\bm{r}_{\alpha} - \frac{2}{5}\bm{s}_i/2)^2}, \\
\phi_{\beta}(-\bm{s}_i) & = \left( \frac{1}{\pi b^2}\right)^{3/4} e^{-\frac{1}{2b^2} (\bm{r}_{\beta} + \frac{3}{5}\bm{s}_i/2)^2}.
\end{align}
The $\bm{s}_i$, $i=1,2,\cdots,n$, are the generating coordinates, which are introduced to expand the relative motion wave function~\cite{Ping:1998si,Wu:1998wu,Pang:2001xx}. The mixing parameter $\epsilon(\bm{s}_i)$ is variationally determined by the dynamics of the multiquark system.
This procedure allows the multiquark system to choose its favorable configuration with the interactions, and was used to explain the cross-over transition between the hadron phase and the quark-gluon plasma phase~\cite{Mingmei:2007jx}. To test the sensitivity of the model parameters, three sets of parameters, labeled as QDCSM1, QDCSM2, and QDCSM3, are used in the calculations as listed in Table~\ref{parameters}. The set in QDCSM1 is taken from the work of dibaryons~\cite{Huang:2015yza}, the one in QDCSM2 is obtained by fitting the spectra of ground state mesons, and the one in QDCSM3 is obtained by fitting the spectra of ground-state baryons and mesons.

\begin{table}[ht]
\caption{Three sets of model parameters discussed in this work:
$m_{\pi}=0.7~{\rm fm}^{-1}$, $m_{K}=2.51~{\rm fm}^{-1}$,
$m_{\eta}=2.77~{\rm fm}^{-1}$, $\Lambda_{\pi}=4.2~{\rm fm}^{-1}$,
$\Lambda_{K}=5.2~{\rm fm}^{-1}$, $\Lambda_{\eta}=5.2~{\rm
fm}^{-1}$, $\alpha_{\rm ch}=0.027$.\label{parameters}}
\begin{tabular}{lccc}
\hline\hline
~~~~~~~~~~~~~~~~~~~~~~~~ &  ~~~~~~{\rm QDCSM1}~~~~~~ & ~~~~~~{\rm QDCSM2}~~~~~~ & ~~~~~~{\rm QDCSM3}~~~~~~    \\
\hline
$m_{u}/{\rm MeV}$          &    313    & 313    &   313   \\
$m_{s}/{\rm MeV}$          &    573    & 559    &   608   \\
$b /{\rm fm}$             &    0.518  & 0.518  &   0.518  \\
$ a_c/{\rm MeV\,fm}^{-2}$  &    58.0   & 55.9   &   41.1   \\
$ V_0/{\rm fm}^{2}$        &   -1.29   & -0.49  &   -0.93  \\
$\alpha_{s}(\mu_0)$         &   0.51    & 0.92   &   1.60   \\
$\Lambda_{0} /{\rm MeV}$   &   300     & 353    &   217    \\
$\mu_{0} /{\rm MeV}$       &   445.81  & 441.80 &   416.30 \\
\hline\hline
\end{tabular}
\end{table}

\section{Hidden strange baryon-meson system in the QDCSM \label{sec3}}

In this section, we calculate the hidden strange light baryon-meson system with isospin $I=\frac{1}{2}$ and $J^{P}=\frac{1}{2}^{\pm}$, $\frac{3}{2}^{\pm}$, and $\frac{5}{2}^{\pm}$ of $S$, $P$, and $D$ partial waves in the framework of RGM~\cite{Kamimura1977}. As listed in Table~\ref{channels}, the channel coupling effects are taken into account.

\begin{table}
\caption{The coupling channels of each quantum number.\label{channels}}
\begin{tabular}{lcccccccccc}
\hline \hline
 $J^{P}$~~~~~~ & ~~~~~~$^{2S+1}L_{J}$~~~~~~ &  Channels  \\  
\hline
 $\frac{1}{2}^{-}$ & $^{2}S_{\frac{1}{2}}$ & $N\eta'$, $N\phi$, $\Lambda K$, $\Lambda K^{*}$, $\Sigma K$, $\Sigma K^{*}$, $\Sigma^{*}K^{*}$    \\
& $^{4}D_{\frac{1}{2}}$ & $N\phi$, $\Lambda K^{*}$, $\Sigma K^{*}$, $\Sigma^{*}K$, $\Sigma^{*} K^{*}$    \\
\hline
 $\frac{3}{2}^{-}$ & $^{2}D_{\frac{3}{2}}$ & $N\eta'$, $N\phi$, $\Lambda K$, $\Lambda K^{*}$, $\Sigma K$, $\Sigma K^{*}$, $\Sigma^{*}K^{*}$    \\
& $^{4}S_{\frac{3}{2}}$($^{4}D_{\frac{3}{2}}$) & $N\phi$, $\Lambda K^{*}$, $\Sigma K^{*}$, $\Sigma^{*}K$, $\Sigma^{*} K^{*}$    \\
\hline
 $\frac{5}{2}^{-}$ & $^{2}D_{\frac{5}{2}}$ & $N\eta'$, $N\phi$, $\Lambda K$, $\Lambda K^{*}$, $\Sigma K$, $\Sigma K^{*}$, $\Sigma^{*}K^{*}$    \\
& $^{4}D_{\frac{5}{2}}$ & $N\phi$, $\Lambda K^{*}$, $\Sigma K^{*}$, $\Sigma^{*}K$, $\Sigma^{*} K^{*}$    \\
\hline
 $\frac{1}{2}^{+}$ & $^{2}P_{\frac{1}{2}}$ & $N\eta'$, $N\phi$, $\Lambda K$, $\Lambda K^{*}$, $\Sigma K$, $\Sigma K^{*}$, $\Sigma^{*}K^{*}$    \\
& $^{4}P_{\frac{1}{2}}$ & $N\phi$, $\Lambda K^{*}$, $\Sigma K^{*}$, $\Sigma^{*}K$, $\Sigma^{*} K^{*}$    \\
\hline
 $\frac{3}{2}^{+}$ & $^{2}P_{\frac{3}{2}}$ & $N\eta'$, $N\phi$, $\Lambda K$, $\Lambda K^{*}$, $\Sigma K$, $\Sigma K^{*}$, $\Sigma^{*}K^{*}$    \\
& $^{4}P_{\frac{3}{2}}$ & $N\phi$, $\Lambda K^{*}$, $\Sigma K^{*}$, $\Sigma^{*}K$, $\Sigma^{*} K^{*}$    \\
\hline
 $\frac{5}{2}^{+}$ & $^{4}P_{\frac{5}{2}}$ & $N\phi$, $\Lambda K^{*}$, $\Sigma K^{*}$, $\Sigma^{*}K$, $\Sigma^{*} K^{*}$    \\
 \hline\hline
\end{tabular}
\end{table}

To investigate if the baryon-meson bound state can be formed, the resonating-group equation has to be solved. We expand the relative motion wave function between two clusters on Gaussian bases, and then the integro-differential resonating-group equation reduces to an algebraic eigenequation with the energy of the system as the eigenvalues. Practically, the baryon-meson separation is restricted to no greater than 6\,fm to keep the dimension of the matrix manageable. We find none of the $P$-wave or $D$-wave states in Table~\ref{channels} can form a bound state, even if the channel coupling effect is taken into account. The reason is that the interaction of the $P$-wave and $D$-wave channels is repulsive, and thus leads to the energies above the threshold. For the $S$-wave channels, the bound state is solved with the binding energies and the masses of each individual channel and of all coupled channels, as shown in Table~\ref{bound}. We need to mention that the mass of the bound state can be generally splitted into three terms as the baryon mass $M_{\rm baryon}$, the meson mass $M_{\rm meson}$, and the binding energy from interactions $M_{\rm int}$. 
To minimize the theoretical deviations, the former two terms, $M_{\rm baryon}$ and $M_{\rm meson}$, are shifted to the experimental values in~\cite{Olive:2016xmw}.

\begin{table}
\caption{The binding energy and the total energy of each individual channel and all coupled channels for the two $S$-wave bound states with the quantum numbers $J^P=\frac{1}{2}^-$ and $\frac{3}{2}^-$. The values are provided in units of MeV, and ``ub'' represents unbound.\label{bound}}
\begin{tabular}{lcccccc}
\hline\hline
Channel & \multicolumn{3}{c}{$J^{P}=\frac{1}{2}^{-}$} & \multicolumn{3}{c}{$J^{P}=\frac{3}{2}^{-}$} \\ 
 \cline{2-4} \cline{5-7}
 & ~~~QDCSM1~~~ & ~~~QDCSM2~~~ & ~~~QDCSM3~~~ & ~~~QDCSM1~~~ & ~~~QDCSM2~~~ & ~~QDCSM3~~  \\
\hline
 $N\eta'$ &      ub      &      ub      &      ub      &     ---      &     ---      &     ---       \\
 $N\phi$ &      ub      &      ub      &      ub      &      ub      &      ub      &      ub       \\
 $\Lambda K$ &  ub      &      ub      &      ub      &     ---      &     ---      &     ---       \\
 $\Lambda K^{* }$ & ub   &      ub      &      ub      &      ub      &      ub      &      ub       \\
 $\Sigma K$ & $-6.7/1681.3$ & $-26.8/1661.2$ & $-4.9/1683.1$ &    ---      &     ---      &     ---       \\
 $\Sigma K^{*}$ & $-8.9/2076.1$ & $-30.6/2054.4$ & $-22.4/2062.2$ & $-21.6/2063.4$ & $-21.1/2063.9$ & $-21.2/2063.8$ \\
 $\Sigma^{*}K$ &  ---      &     ---      &     ---    & $-10.4/1869.6$ & $-15.5/1864.5$ & $-11.1/1868.9$  \\
 $\Sigma^{*}K^{*}$ & $-17.3/2259.7$ & $-87.0/2190.0$ & $-73.9/2203.1$ & $-11.3/2265.7$ & $-18.4/2258.6$ & $-27.2/2249.8$ \\
 coupled & $-16.0/1881.0$ & $-20.0/1877.0$ & $-24.3/1872.7$ & $-10.1/1948.9$ & $-7.7/1951.3$ & $-1.6/1957.4$ \\
\hline\hline
\end{tabular}
\end{table}

For single-channel calculations, neither $\Lambda K$ or $\Lambda K^{*}$ is bound. This agrees with the repulsive nature of the interaction between $\Lambda$ and $K$ (or $K^*$). The attractions of $N\eta'$ and $N\phi$ channels are too weak to bind the two particles, while the strong attractive interaction between $\Sigma$ (or $\Sigma^*$) and $K$ (or $K^*$) leads to the total energy below the threshold of the two particles. However, including the channel-coupling effect, we find a $J^P=\frac{1}{2}^-$ bound state with $N\eta'$ as the main component and a $J^P=\frac{3}{2}^-$ bound state with $N\phi$ as the main component. Therefore the channel-coupling effect plays an important role in the quark model calculation of the baryon-meson bound state. With the three sets of model parameters, the mass of the $\eta'-N$ bound state varies from $1872.7$\,MeV to $1881.0$\,MeV, and the mass of the $\phi-N$ bound state varies from $1948.9$\,MeV to $1957.4$\,MeV.

\begin{figure}
\begin{center}
\includegraphics[width=0.7\textwidth]{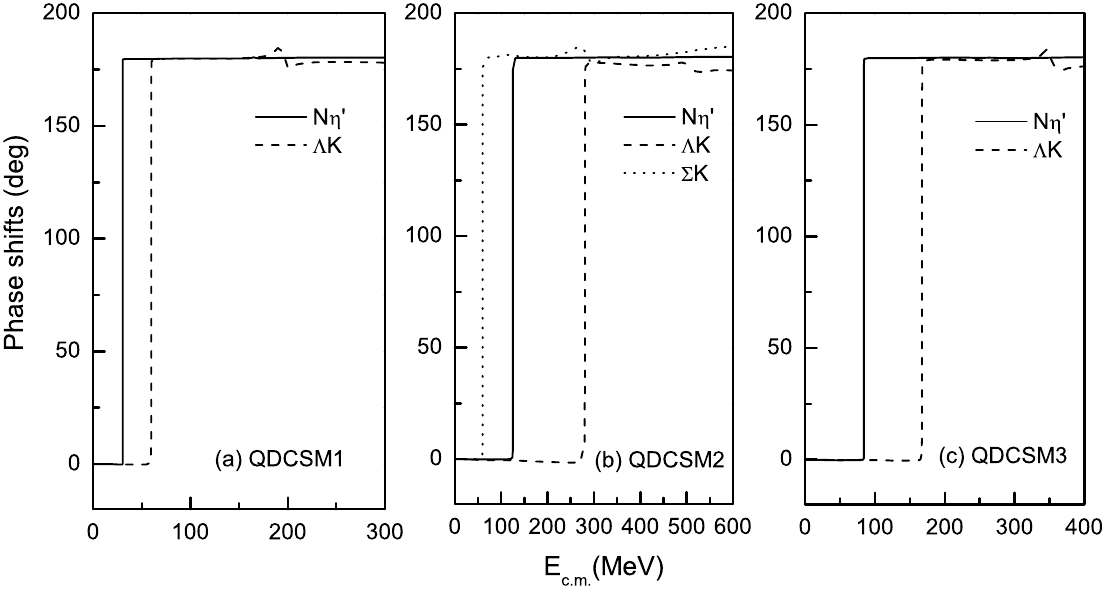}
\caption{The phase shifts of different scattering channels for the $J^{P}=\frac{3}{2}^{-}$ systems.\label{phaseshift}}
\end{center}
\end{figure}

For the main purpose of this paper, we only focus on the $\phi-N$ bound state from now on. Because it has the same quantum number of $N$ baryon states, we label it as $N_{s\bar{s}}$ to indicate its hidden strange structure. To obtain the decay width of $N_{s\bar{s}}$, we calculate the phase shifts of various possible scattering channels which couple to the $S$-wave $J^P=\frac{3}{2}^-$ channels. The phase shift results are shown in Fig.~\ref{phaseshift}. The coupling to the $^2D_{\frac{3}{2}}$ open channel changes the $N_{s\bar{s}}$ bound state into an elastic resonance with the phase shifted by $\pi$ at the resonance mass. The two-body decay channels are $N\eta'$, $\Lambda K$, and $\Sigma K$. We need to note that the theoretical kaon mass in QDCSM1 and QDCSM3 is larger than the real mass, and results in the total energy of $\Sigma K$ being above the $N_{s\bar{s}}$ mass. Thus the $\Sigma K$ decay channel is forbidden with these two parameter sets.

The mass of $N_{s\bar{s}}$ bound state can be obtained from the resonance phase shift point via the relation,
\begin{equation}
M=M_{\rm baryon} + M_{\rm meson} + E_{\rm c.m.},
\end{equation}
where $E_{\rm c.m.}$ is the incoming kinetic energy of the two scattering particles in the center-of-mass frame, and the other two terms, $M_{\rm baryon}$ and $M_{\rm meson}$, are shifted to the experimental values~\cite{Olive:2016xmw}. As shown in Table~\ref{mass}, the resonance mass values from different scattering channels are equal to each other within the numerical precision. We should point out that the small difference between the mass value from the resonance scattering approach and the one from the Hamiltonian eigenequations is from the tensor coupling.

\begin{table}
\caption{The $N_{s\bar{s}}$ bound state mass calculated from the $^{2}D_{\frac{3}{2}}$ scattering channels. The values are provided in units of MeV.\label{mass}}
\begin{tabular}{lccccccc}
\hline\hline
 Scattering channel &  ~~~~~~{\rm QDCSM1}~~~~~~ & ~~~~~~{\rm QDCSM2}~~~~~~ & ~~~~~~{\rm QDCSM3}~~~~~~  \\   
\hline
 ~$N\eta'$~~~ & $1947.998$ & $1949.485$  & $1955.988$ \\
 ~$\Lambda K$~~~ & $1947.975$ & $1949.480$ & $1955.910$ \\
 ~$\Sigma K$~~~ & -- & $1949.597$ & -- & \\
\hline\hline
\end{tabular}
\end{table}

The phase shifts in Fig.~\ref{phaseshift} show a narrow resonance, and the decay width of the $N_{s\bar{s}}$ to each channel is obtained from the shape of the resonance. The small width is from the tensor coupling. This is similar to the case of $N-\Omega$ state, which is also a narrow resonance in the $D$-wave $\Lambda\Xi$ scattering process~\cite{Chen:2011zzb,Huang:2015yza}.

Apart from the decay channels calculated above, the decay of $N_{s\bar{s}}$ bound state can also be caused by the $\phi$ meson decay in the system, and the partial width from the $\phi$ meson decay is not negligible. Following the procedure in~\cite{JuliaDiaz:2007kz}, the nucleon-bound $\phi$ meson decay width is related to the free $\phi$ meson decay width. Because the charged kaon and the neutral kaon are not differentiated in our model, the total width of $\phi$ meson decay induced channels are calculated by dividing the partial width of the $K^+K^-$ channel by the average branch ratio of the $K^+K^-$ and the $K^0_LK^0_S$ channels. We point out that the width of other channels, such as $\rho\pi^0$ and $3\pi$, is effectively included after dividing the partial width of the $KK$ channel by its branch ratio. In Table~\ref{width}, we summarize $N_{s\bar{s}}$ decay widths and branch ratios of each channel, and the one induced by $\phi$ meson decays is dominant in our model calculation.

\begin{table}
\caption{The decay widths and branch ratios of each decay channel of $N_{s\bar{s}}$ bound state.\label{width}}
\begin{tabular}{lccccccc}
\hline\hline
Decay channel~~~ & \multicolumn{2}{c}{~~~~~~~~~QDCSM1~~~~~~~~~~~~}&\multicolumn{2}{c}{~~~~~~~~~QDCSM2~~~~~~~~~~~~} &\multicolumn{2}{c}{~~~~~~~~~QDCSM3~~~~~~~~~~~~} \\ 
\cline{2-3} \cline{4-5} \cline{6-7}
& $\Gamma_{i} ({\rm MeV})$ & $\Gamma_{i}/\Gamma (\%)$ & $\Gamma_{i} ({\rm MeV})$ & $\Gamma_{i}/\Gamma (\%)$ & $\Gamma_{i} ({\rm MeV})$ & $\Gamma_{i}/\Gamma (\%)$ \\
\hline
 ~$N\eta'$ & $0.002$ & $0.1$  & $0.022$ & $0.5$ & $0.009$ & $0.2$ \\
 ~$\Lambda K$ & $0.011$ & $0.3$ & $0.120$ & $2.9$  & $0.055$ & $1.2$ \\
 ~$\Sigma K$ & -- & $0.0$ & $0.060$ & $1.5$ & -- & $0.0$ \\
 ~$\phi$ decays & $3.619$ & $99.6$ & $3.892$ & $95.1$ & $4.616$ & $98.6$ \\
\hline\hline
\end{tabular}
\end{table}

\section{The \texorpdfstring{$\phi-N$}{phi-N} bound state production on a nuclear target \label{sec4}}

It is pointed out in~\cite{Gao:2000az} that the subthreshold production of $\phi$ meson inside a nuclear medium will enhance the probability of the formation of the $\phi-N$ bound state. In this section, we take a gold target as an example to study the production of the $N_{s\bar{s}}$ bound state, and show the feasibility of the experimental search by simulation. For simplicity, we will only present the results with the parameter set of QDCSM2, and similar results are expected with the other two parameter sets.

\begin{figure}[ht]
\includegraphics[width=0.5\textwidth]{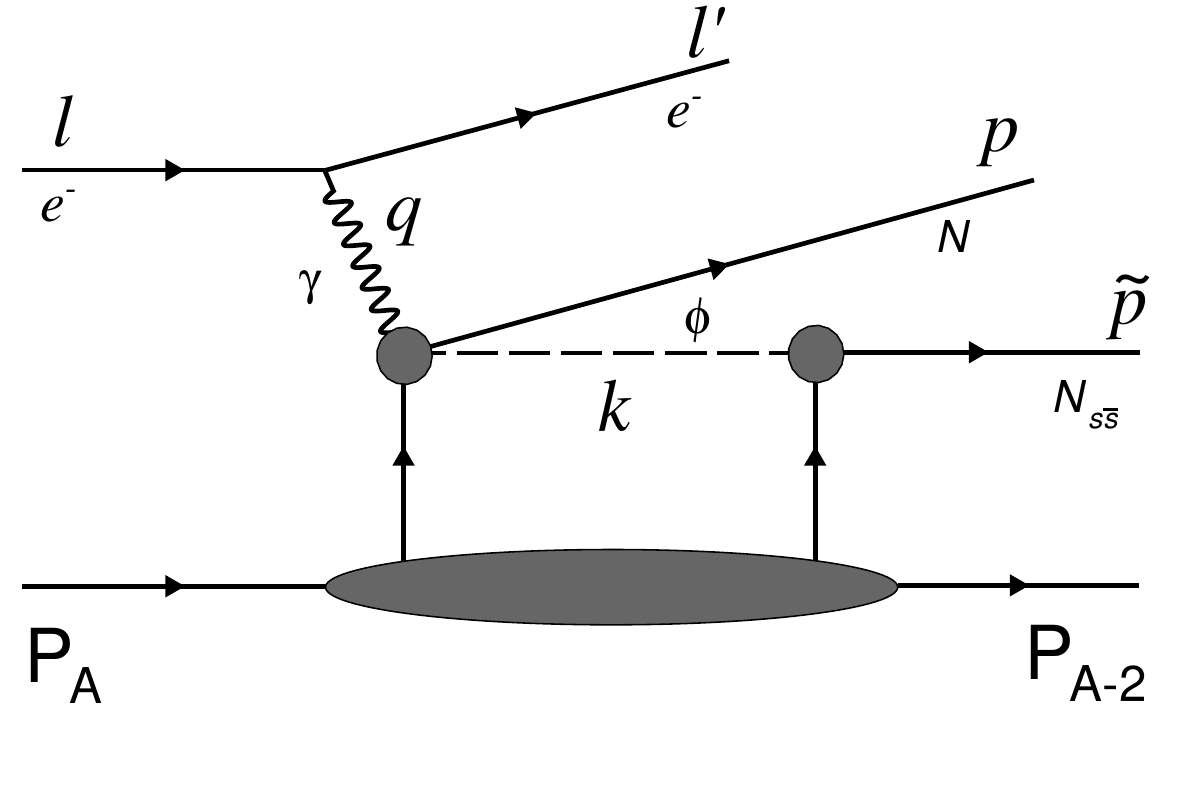}
\caption{The mechanism of $N_{s\bar{s}}$ bound state electroproduction on a nuclear target.\label{feyndiag}}
\end{figure}

\begin{figure}[ht]
\includegraphics[width=0.45\textwidth]{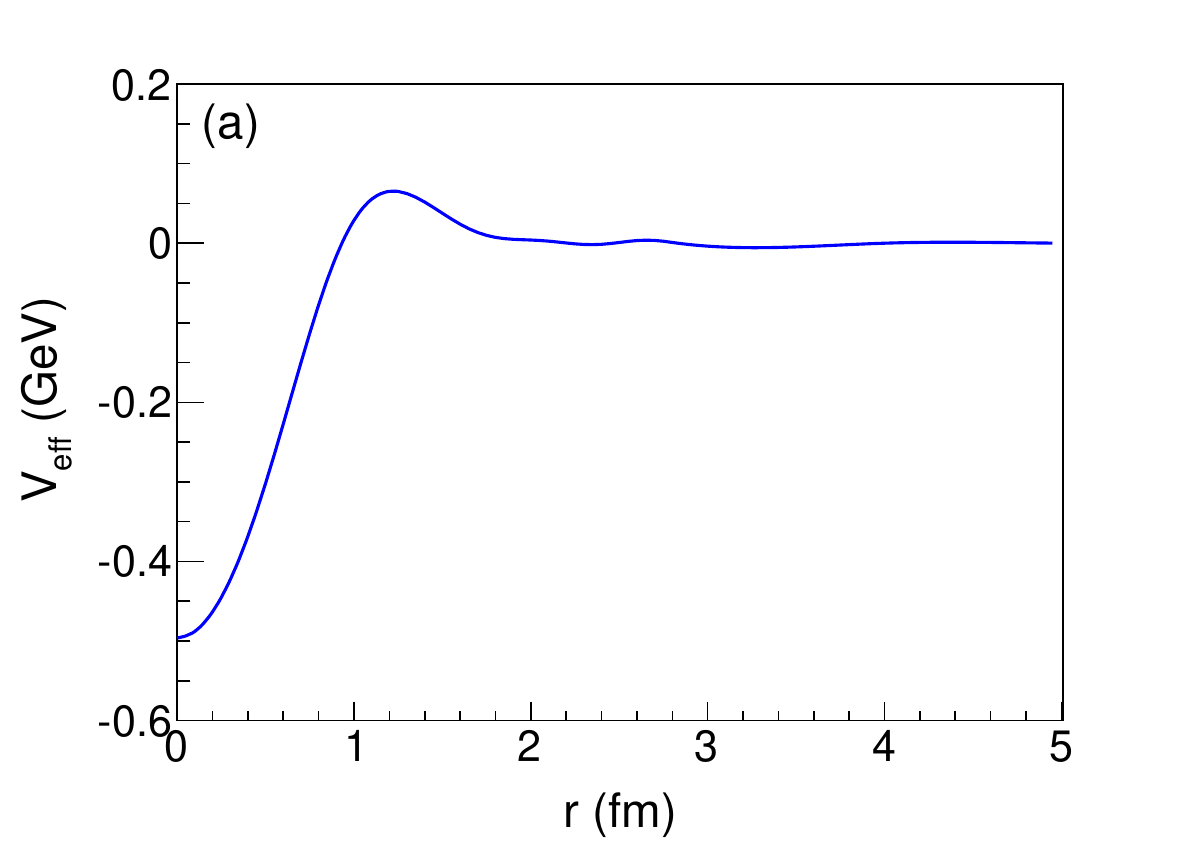}
\includegraphics[width=0.45\textwidth]{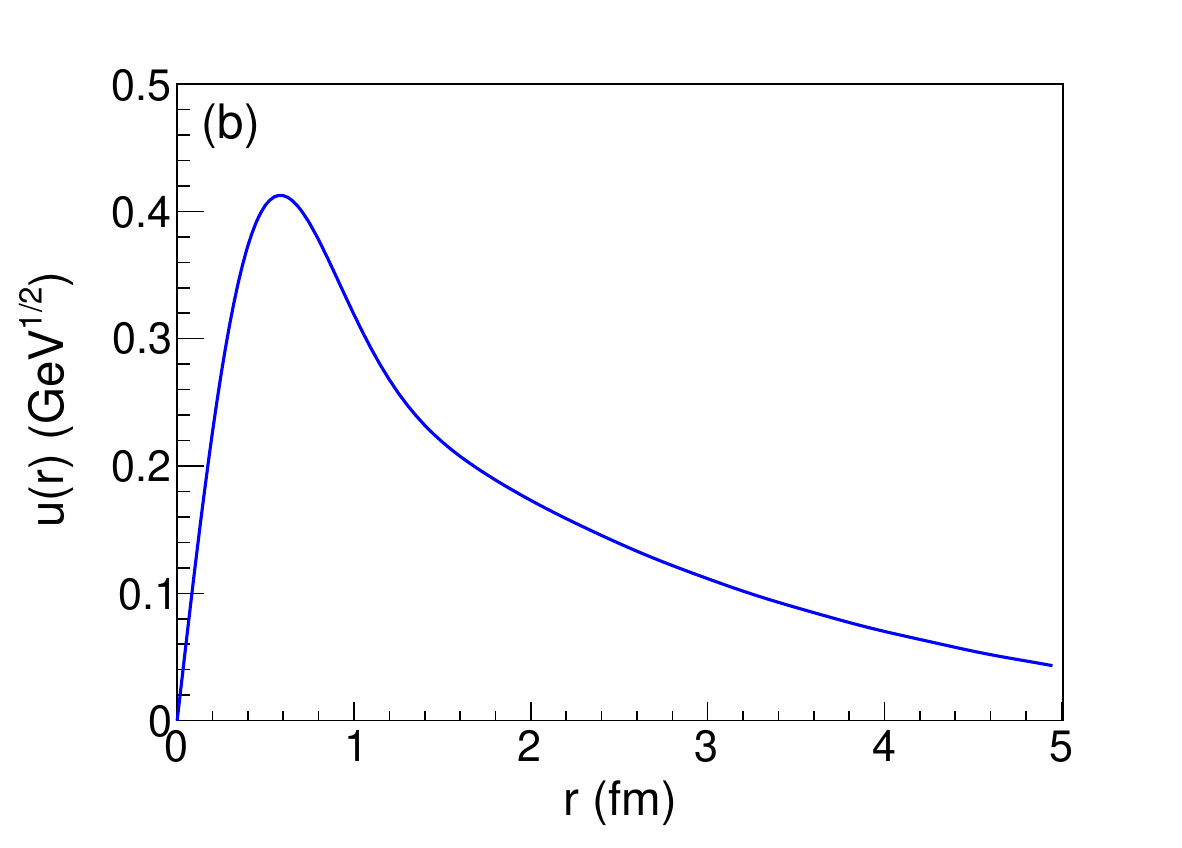}
\caption{(color online) (a): The effective potential between $\phi$ and $N$. (b): The radial wave function $u(r)=rR(r)$ of the bound state $N_{s\bar{s}}$.\label{Veff_wf}}
\end{figure}

\begin{figure}[ht]
\includegraphics[width=0.5\textwidth]{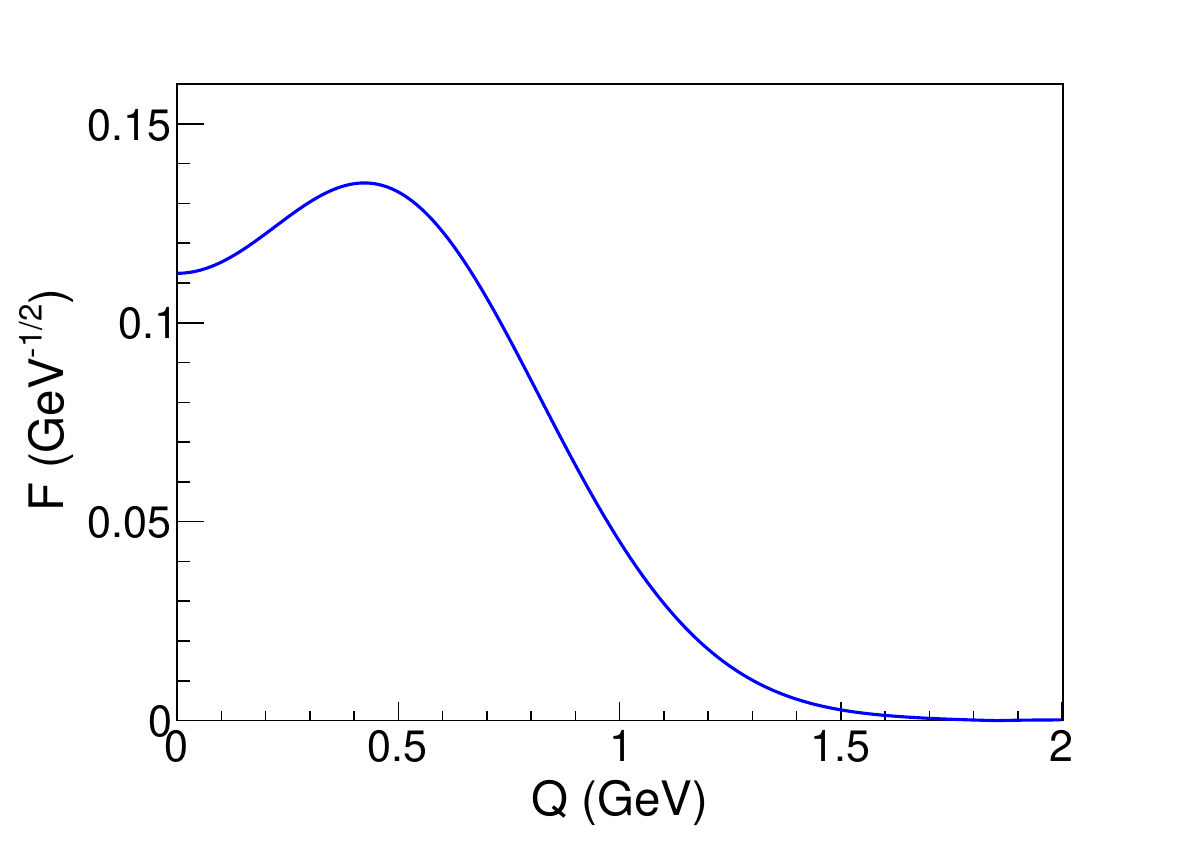}
\caption{(color online) The amplitude $F(Q)$ of the $N_{s\bar{s}}$ bound state formation. The $Q$ is the relative momentum of the $N\phi$ system. \label{FQ}}
\end{figure}

As illustrated in Fig.~\ref{feyndiag}, the reaction takes place in two steps. First the $\phi$ meson is produced from a nucleon in a nuclear medium, and then it interacts with another nucleon around to form the bound state $N_{s\bar{s}}$. The amplitude of the formation of the bound state can be calculated from the effective potential $V_{\rm eff}(r)$ and the radial wave function $R(r)$ as
\begin{equation}
F(Q)=\langle N_{s\bar{s}}|V_{\rm eff}|\phi(\bm{Q}),N(-\bm{Q})\rangle=\frac{\sqrt{4\pi}}{(2\pi)^{3/2}}\int\frac{\sin(Q\,r)}{Q\,r}V_{\rm eff}(r)R(r)r^2dr,
\end{equation}
where $Q$ is the incoming momentum of the $\phi$ meson in the center-of-mass frame of the $N\phi$ system. Using the effective potential and the wave function in Fig.~\ref{Veff_wf}, which are calculated in the QDCSM, we get the amplitude result as shown in Fig.~\ref{FQ}.

For the $\phi$ meson production, the amplitude is extracted from the $\phi$ meson near threshold photoproduction differential cross section data in~\cite{Dey:2014tfa} via
\begin{equation}
\frac{d\sigma}{d\cos\theta}=\frac{|\mathcal{M}|^2}{32\pi q_c(E_N(q_c)+q_c)}\frac{Q_c}{E_\phi(Q_c)+E_N(Q_c)},
\end{equation}
where $\mathcal{M}$ is the invariant amplitude, $q_c$ is the relative momentum of the incoming $\gamma N$ system, and $Q_c$ is the relative momentum of the outgoing $\phi N$ system. The $\theta$ represents the polar angle of the produced $\phi$ meson in the center-of-mass frame with respect to the direction of the incoming photon. The spin dependent effects are neglected here. Namely the extracted amplitude is spin averaged.

\begin{figure}[ht]
\includegraphics[width=0.48\textwidth]{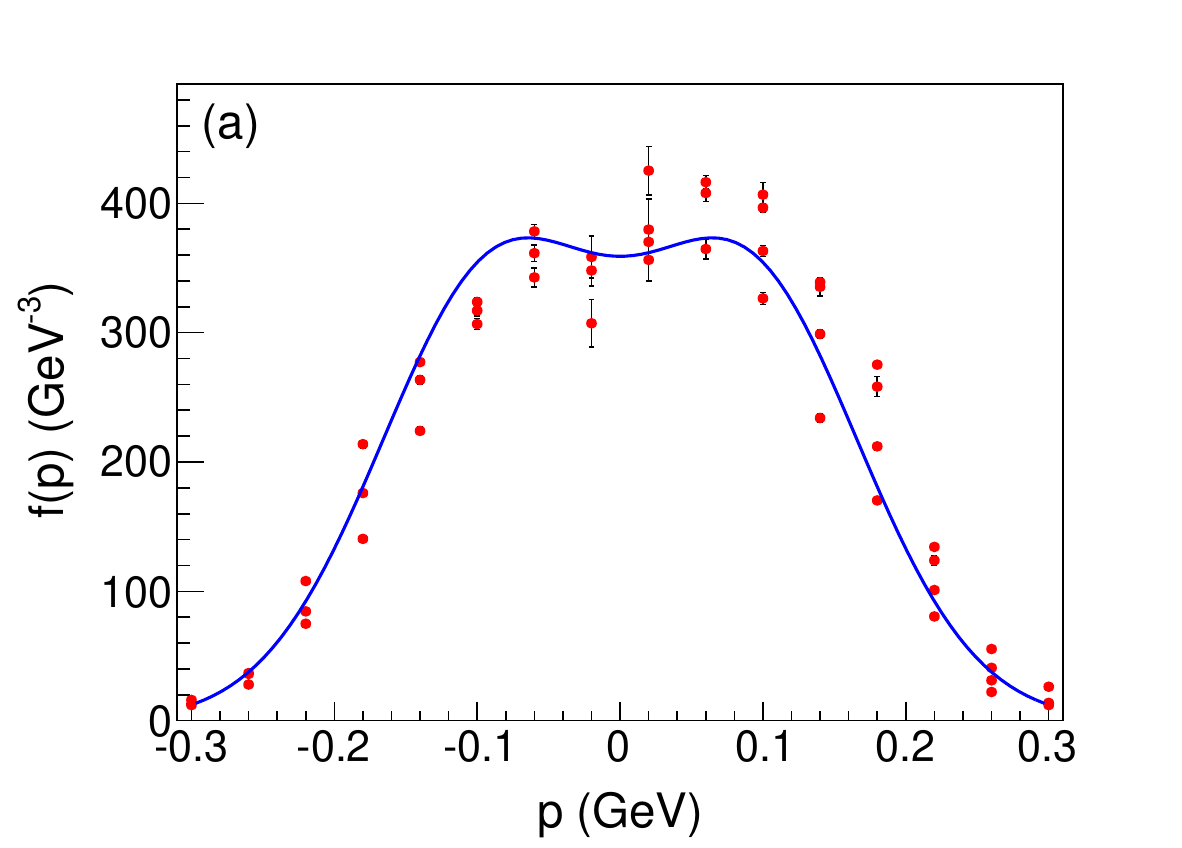}
\includegraphics[width=0.48\textwidth]{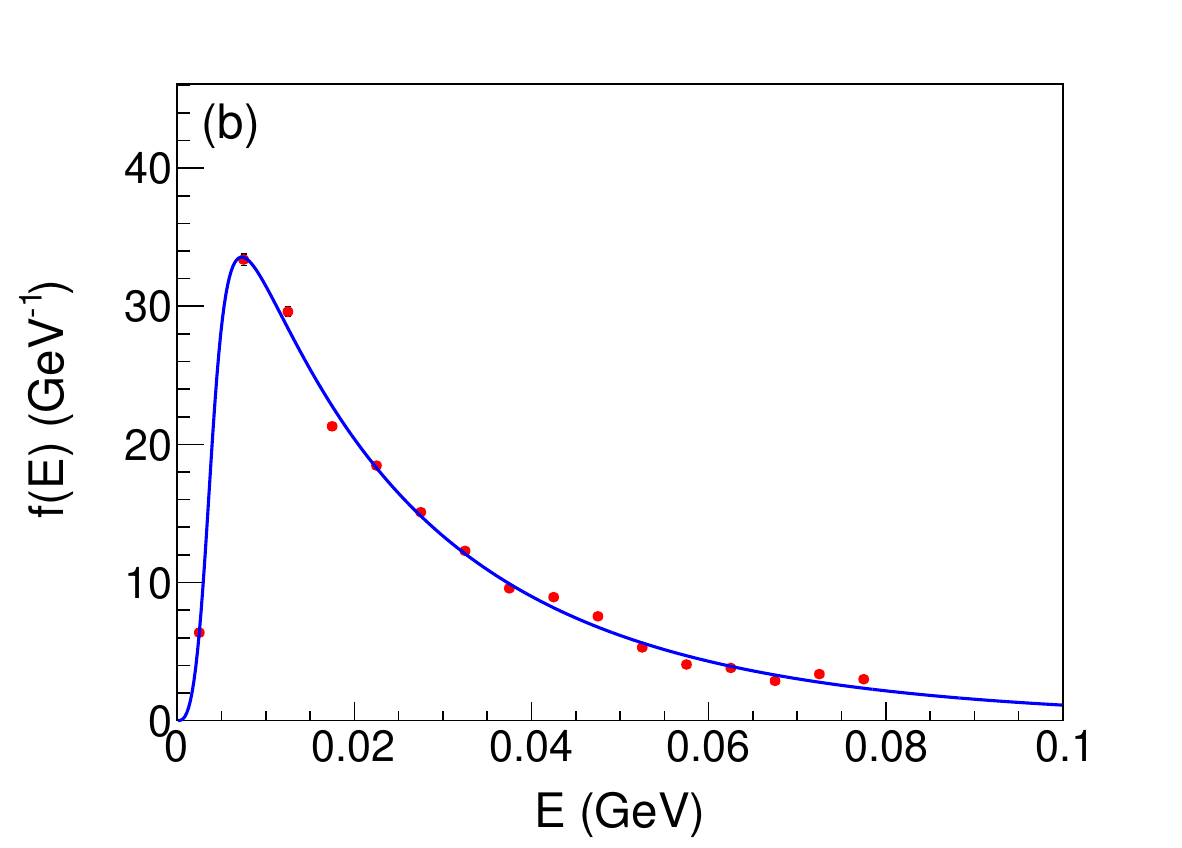}
\caption{(color online) The momentum, shown in (a), and the missing energy, shown in (b), distributions of nucleons inside a gold nucleus. The data are taken from Ref.~\cite{Dutta:2003yt}, and the curves are the extracted distributions we use in the calculation and the simulation.\label{goldnucleus}}
\end{figure}

Because of the momentum mismatch, it is not likely to find the $N_{s\bar{s}}$ bound state in the $\phi$ production from a proton target. With the help of the Fermi motion, the probability of the bound state formation is expected to be enhanced in the $\phi$ meson sub- or near-threshold production from heavy nuclei. Therefore we choose gold ($^{197}$Au) as the target here. To describe bound nucleons in a gold nucleus, we extract the momentum and the energy distributions of nucleons inside a gold nucleus from the data in~\cite{Dutta:2003yt}. Here we assume an isotropic momentum distribution. The results are shown in Fig.~\ref{goldnucleus}.

\begin{figure}[ht]
\includegraphics[width=0.5\textwidth]{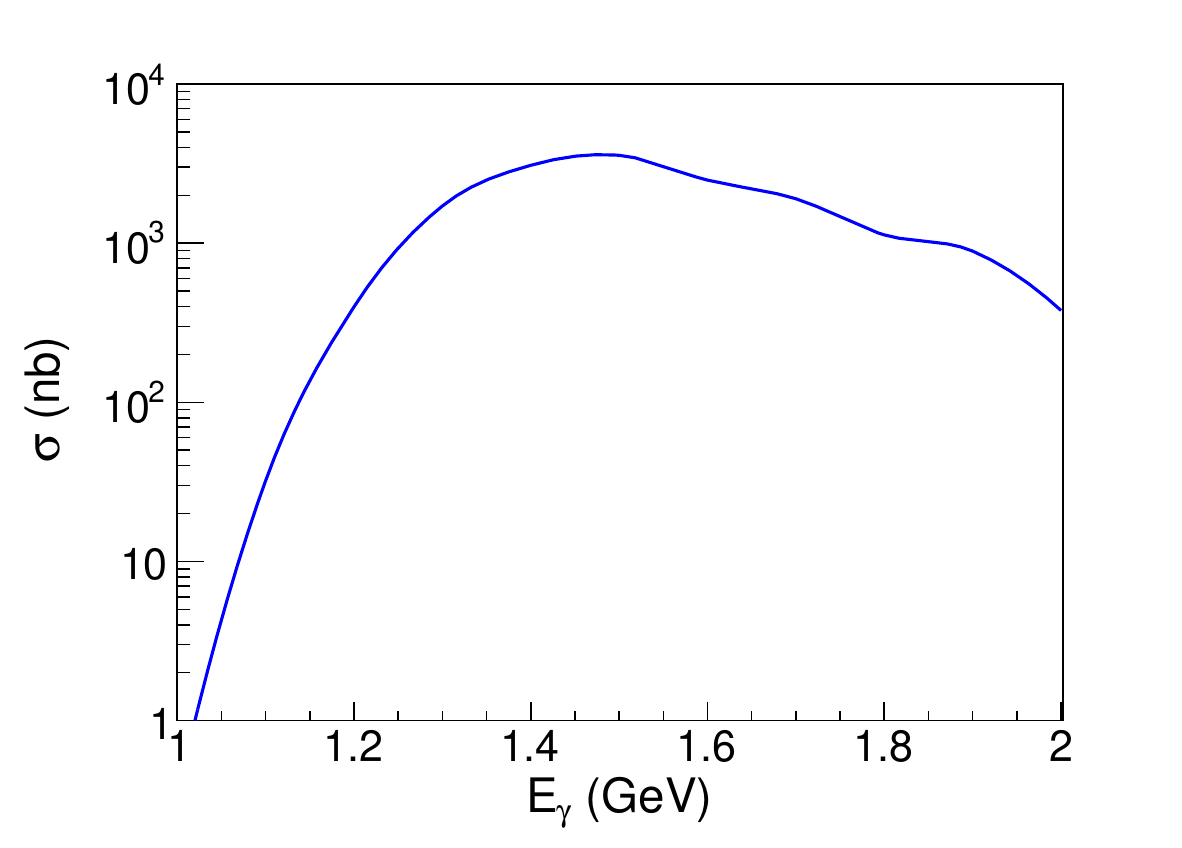}
\caption{(color online) The cross section of the $N_{s\bar{s}}$ bound state photoproduction on a gold target.\label{sigmaphoton}}
\end{figure}

Following the procedure in~\cite{Gao:2000az}, we calculate the total cross section of the $N_{s\bar{s}}$ bound state photoproduction on a gold target. As shown in Fig.~\ref{sigmaphoton}, the cross section of $N_{s\bar{s}}$ photoproduction has a maximum below the threshold $E_\gamma=1.57$\,GeV. This feature is consistent with the calculation in~\cite{Gao:2000az}. As expected, the cross section drops with the photon energy above the threshold because of the increasing $\phi N$ relative momentum.

To investigate the feasibility of an experimental search for the $N_{s\bar{s}}$ bound state, we perform a simulation according to the configuration of CLAS12 at Jefferson Lab. Based on the calculation in the previous section, the $N_{s\bar{s}}$ decay is dominated by the $\phi$ meson decay. Thus the bound state can be reconstructed from the $NKK$ channel. Because of the detection issue, we restrict the nucleon in the bound state to be a proton, which means the cross section is divided by a factor of $Z/A=79/197$, because the proton and the neutron are not differentiated in our model calculation. In addition, the two kaons in the final state are required to be $K^+K^-$ because of the difficulty of the $K^0_L$ detection. Assuming the same branch ratio of the $K^+K^-$ decay channel of the nucleon-bound $\phi$ meson as the one of the free $\phi$ meson, 48.9\%~\cite{Olive:2016xmw}, we obtain the branch ratio of the $pK^+K^-$ decay channel for $N_{s\bar{s}}$ as 46.5\% (see Table~\ref{width}).

Instead of the photoproduction, we perform the Monte Carlo simulation of the electroproduction with one photon exchange approximation as illustrated in Fig.~\ref{feyndiag}, because photon beams will not be readily available for CLAS12 in Hall B at Jefferson Lab. The forward tagger~\cite{ForwardTagger}, which covers the polar angle $2.5^\circ \sim4.5^\circ$ and the energy above 0.5\,GeV, can be used to detect the scattered electron to select the low $Q^2$ events. Apart from the scattered electron, a triple coincident detection of $pK^+K^-$ in the final state is required to reconstruct the $N_{s\bar{s}}$ bound state.

In addition to the simulation of the signal channel, an estimation of the background is necessary to validate the experimental feasibility. In this study, four background channels are estimated. The first one is from the same reaction as the $N_{s\bar{s}}$ production process in Fig.~\ref{feyndiag}, but the detected proton in the final state is not from the bound state decay. The second one is the $\phi$ meson production process without the formation of the bound state. As mentioned above, the amplitude extracted from the data in~\cite{Dey:2014tfa} is used in the simulation of this channel. The third one is the production of $\Lambda(1520)K^+$ with the $\Lambda(1520)$ decaying into $pK^-$. Similar to the $\phi$ production case, the amplitude of the near threshold production of $\Lambda(1520)K^+$ is extracted from the differential cross section data in~\cite{Moriya:2013hwg}. In this process, the distributions of $K^+$ and $K^-$ in the final state are different. The fourth one is the direct $K^+K^-$ production near the threshold. In this case, we model the cross section by using the amplitude of $\phi$ production but replacing the mass with the invariant mass of the $K^+K^-$ system. Actually the value of this amplitude is not very critical, because we can separate it from the signal as discussed below.

\begin{figure}[ht]
\includegraphics[width=0.495\textwidth]{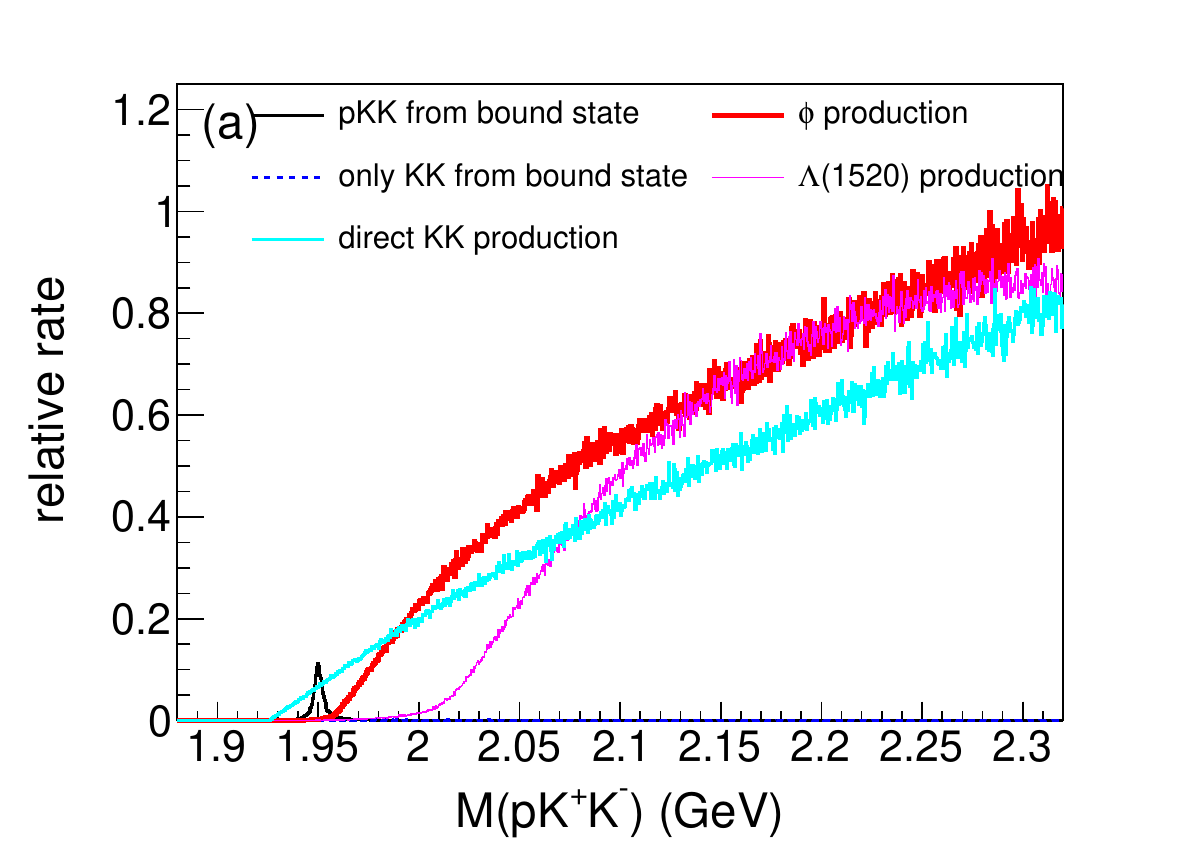}
\includegraphics[width=0.495\textwidth]{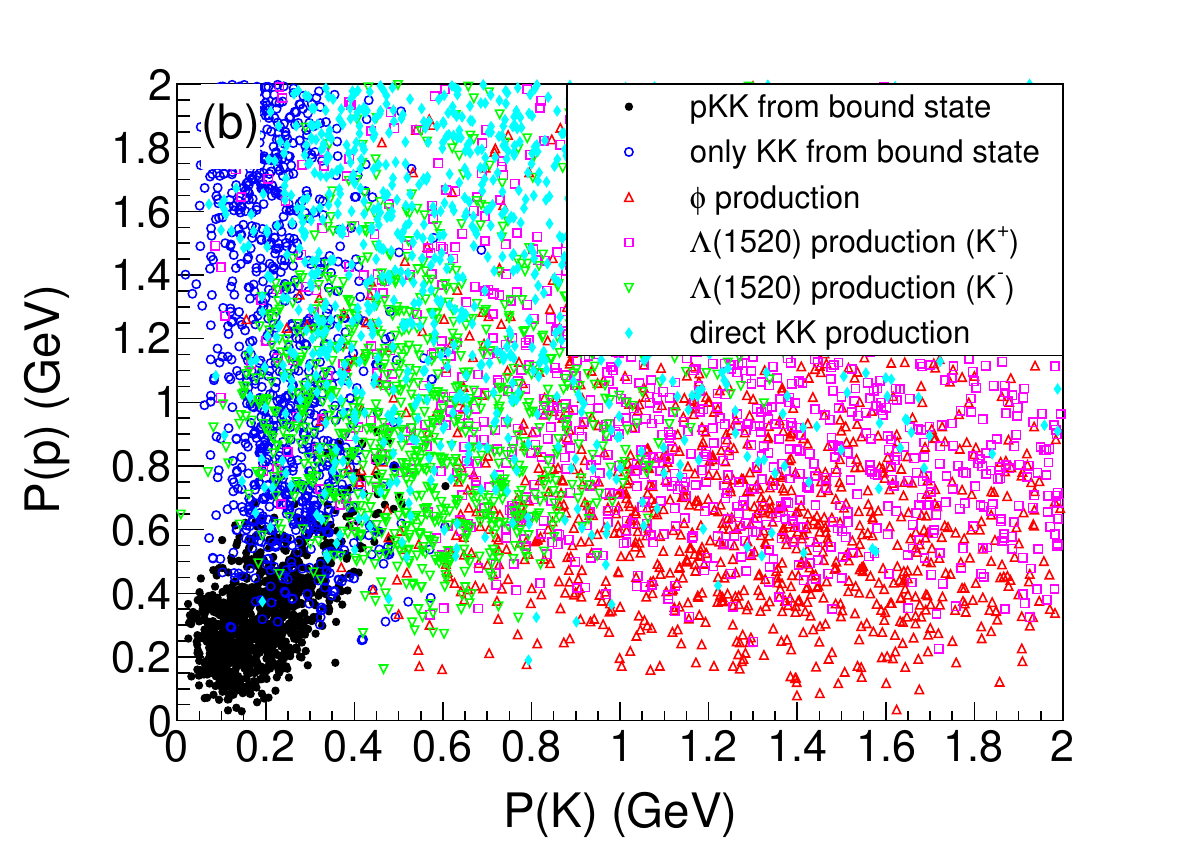}
\caption{(color online) The comparison between the signal and the background channels with the scattered electron detected by the forward tagger. The plot (a) shows the comparison of the relative rate in the invariant mass spectra of the $pK^+K^-$ system. The plot (b) shows the momentum distributions of the proton and kaon produced from each channel. For better visibility, an equal number of points are plotted in (b) for each channel, and the density of the points reflects the distribution. The black (dark) thick curve in (a) and the black solid round points in (b) are from the signal channel with the proton and the two kaons decayed from the $N_{s\bar{s}}$ bound state. The blue (dark gray) dashed curve in (a) and the blue hollow circles in (b) are also from the reaction with $N_{s\bar{s}}$ productions, but the proton is not decayed from the bound state.\label{pKK_vertex}}
\end{figure}

In the Monte Carlo simulation, we choose a 4.4\,GeV electron beam with a 100\,nA beam current and a 0.138\,mm thickness gold target. It corresponds to the luminosity of $10^{35}\,eN$\,cm$^{-2}\cdot$s$^{-1}$. The masses of $\phi$ and $\Lambda(1520)$ are sampled according to the Breit-Wigner distribution. The particles from decays are generated according to the phase space distribution in the center-of-mass frame and then boosted to the laboratory frame according to the four-momentum of the parent particle. The mass and width of the $N_{s\bar{s}}$ bound state is chosen as 1950\,MeV and 4.094\,MeV according to the model calculation results in Tables~\ref{mass} and~\ref{width}. In Fig.~\ref{pKK_vertex}, the invariant mass spectra of the $pK^+K^-$ final state from the signal and the background channels are compared. One can observe that the signal overlaps with the background in the $pK^+K^-$ spectra.

As suggested in~\cite{Liska:2007de}, the signal could be separated from the background by using the information of the proton-kaon momentum correlation. In Fig.~\ref{pKK_vertex}, we show the two-dimensional distribution of the proton and kaon momenta for each channel. It is clearly shown that the signal events are in the low momentum region. Therefore we apply the cuts $p(K^\pm)<350$\,MeV and $p(p)<500$\,MeV to remove the background events with relatively high momenta.

\begin{figure}
\includegraphics[width=0.495\textwidth]{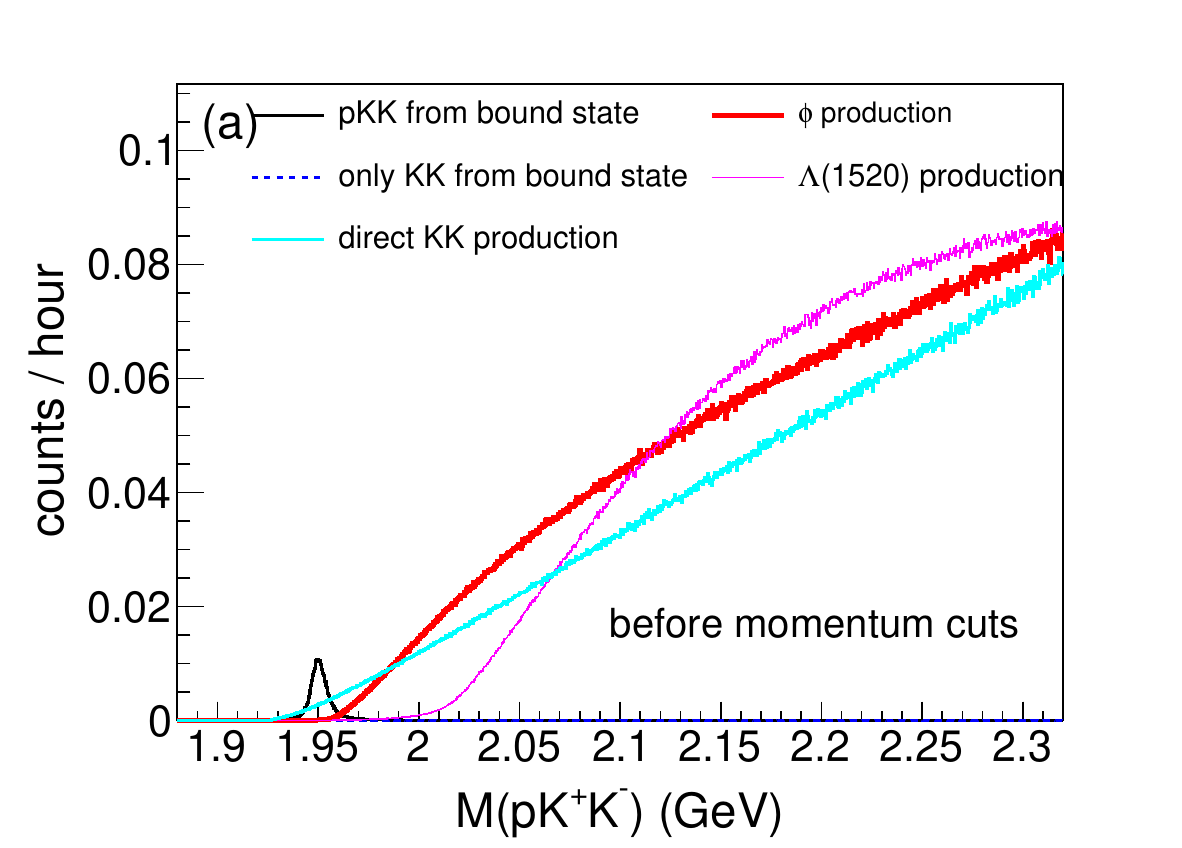}
\includegraphics[width=0.495\textwidth]{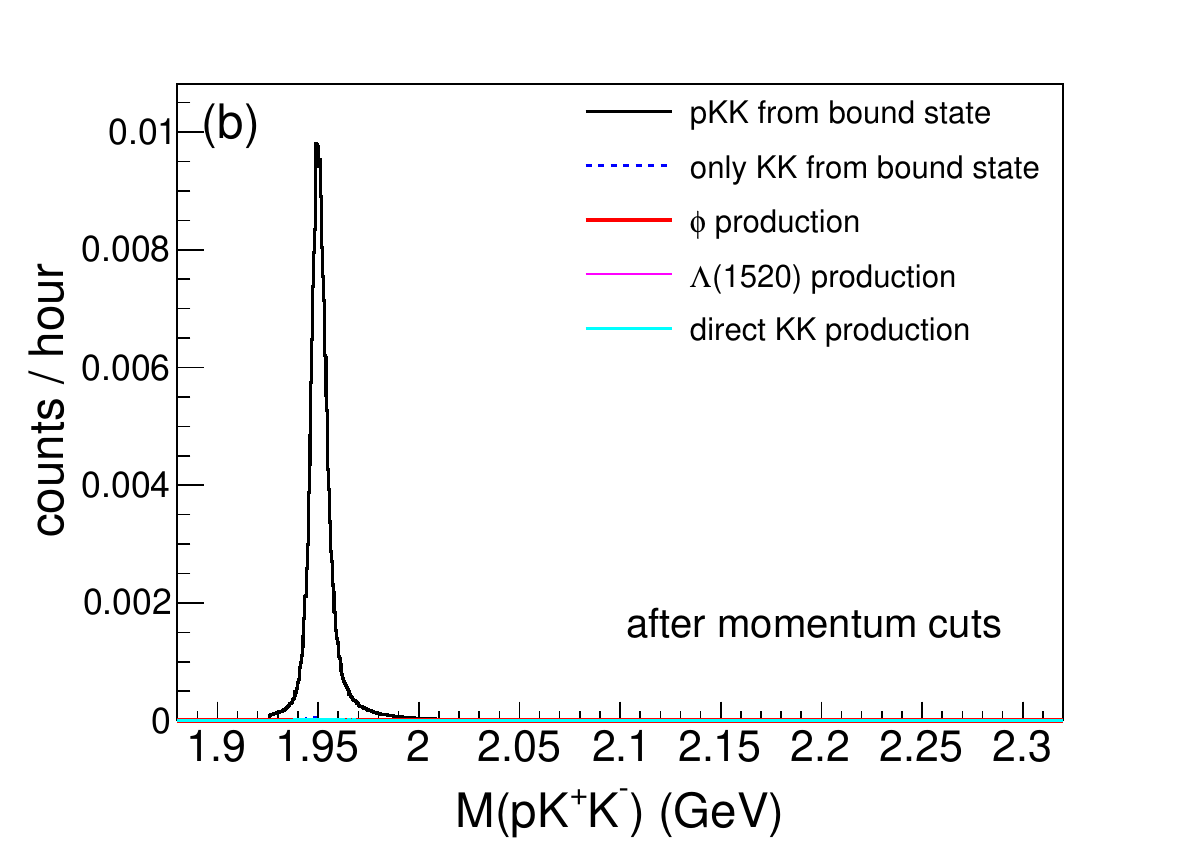}
\caption{(color online) The detected signal and background rates. The plot (a) shows the results before the momentum cuts $p(K)<350$\,MeV and $p(p)<500$\,MeV, and the plot (b) shows those after the momentum cuts. The momentum, the polar angle, and the azimuthal angle of the detected particles are all smeared according to the resolutions of the CLAS12 and the BONUS12 detectors.\label{detected}}
\end{figure}

To improve the detection of the low momentum particles, we propose to use both the CLAS12 and the BONUS12 detectors in Hall B at Jefferson Lab. The BONUS12 is designed to be placed around the target with a polar angle coverage from $20^\circ$ to $160^\circ$ to detect low momentum particles~\cite{BONUS12}. In the simulation, the BONUS12 is set to detect the proton and charged kaons with momenta between 60\,MeV and 200\,MeV. For particles with momenta below 60\,MeV, we assume no detections, and for particles with momenta above 200\,MeV, we leave it to CLAS12 for possible detections. In addition, the energy loss in the target is taken into account based on the stopping power data by NIST~\cite{NIST}, though the effect is negligible when the target is very thin. For CLAS12, the forward detector covers the polar angle from $5^\circ$ to $35^\circ$, and the central detector covers the polar angle from $35^\circ$ to $125^\circ$~\cite{Stepanyan:2010kx}. Because the BONUS12 is close to the target, the $K^\pm$ decay effect is neglected if it is detected by BONUS12. However, a weighting factor is multiplied by assuming a 2\,m traveling distance if the $K^\pm$ is detected by CLAS12.
Concerning the six-fold configuration of the CLAS12 detector, we add a weighting factor of 80\% for each particle detected by CLAS12 to account for the acceptance gaps in the azimuthal angle.
To have more realistic estimation, we also smear the momentum, the polar angle, and the azimuthal angle of the detected particles according to the detector resolutions~\cite{BONUS12,Stepanyan:2010kx}. The results are shown in Fig.~\ref{detected}. It shows that the momentum cuts significantly reduce the background.
However, the signal rate in the region of $1940\,{\rm MeV} < M(pK^+K^-) < 1960\,{\rm MeV}$ drops from 1.64/h before the momentum cuts to 1.50/h after the cuts, {\it i.e.} only 10\% signal events are removed by these cuts. In addition, assuming a total detector efficiency of 50\%, we still have the signal rate as 0.75/h.

Apart from the electroproduction as the example we presented above, the photoproduction can be a complementary approach to search for the bound state. It will be possible if in future the photon beam is available for CLAS12 in Hall B or the flux of the photon beam is enhanced for GlueX in Hall D. Experimental facilities other than those at Jefferson Lab may also find opportunities.

\section{Discussions and conclusions}

In this paper, we investigate the feasibility of the experimental search for the $\phi-N$ bound state $N_{s\bar{s}}$, which is obtained from the calculation in the QDCSM. It can be viewed as a hidden strange pentaquark candidate.

We perform a calculation of the hidden strange light baryon-meson system in the QDCSM. By solving the algebraic resonating-group eigenequation, two bound states are found as one with $J^P=\frac{1}{2}^-$ dominated by the $N\eta'$ component and the other one, labeled as $N_{s\bar{s}}$, with $J^P=\frac{3}{2}^-$ dominated by the $N\phi$ component. Comparing with the single channel calculations, we find that the channel coupling effect in the model is important to determine the existence of the bound state. The decay properties of $N_{s\bar{s}}$ are calculated from the phase shifts in the resonance scattering processes. As $N$ and $\phi$ have no common quarks, a narrow width about $4$\,MeV is obtained for $N_{s\bar{s}}$.

Based on the results of QDCSM, we calculate the $N_{s\bar{s}}$ photoproduction cross section on a gold target. As expected, the cross section decreases with increasing photon energy above the $\phi$ production threshold, because the probability of the bound state formation drops as the relative momentum of the $N\phi$ system increases. Thus we propose to search for the $N_{s\bar{s}}$ bound state in the sub- or near-threshold $\phi$ productions from heavy nuclei.

The feasibility of the experimental search for the $N_{s\bar{s}}$ state is investigated via a simulation using the electron beam at Jefferson Lab and a forward tagger, the CLAS12, and the BONUS12 detectors. As concluded from the model calculation, the $N_{s\bar{s}}$ decay is dominated by the $\phi$ meson decay. The $N_{s\bar{s}}$ state can be reconstructed from the $pK^+K^-$ events. The background channels with the same final state particles are also estimated. We show that the signal events can be separated from the background by applying a set of cuts to select relatively low momentum events. The cuts will significantly reject the background events with a cost of losing only about 10\% of signal events. Assuming a total detector efficiency of 50\%, the signal rate estimated from the simulation is about 
0.75/h.
Therefore it is feasible to search for the $N_{s\bar{s}}$ bound state. If the photon beam is available for CLAS12 in the future, the photoproduction can serve as a complementary approach. Other facilities apart from those at Jefferson Lab may also have the possibility to search for the bound state.
In addition to the background from the $pK^+K^-$ channels, reactions with $p\pi^+\pi^-$ in the final state can be a dominant background source because of the accuracy of particle identifications. A good $K/\pi$ separation is needed to suppress this background. Therefore, more realistic studies on various experimental issues are required to optimize the experimental conditions to search for the bound state.

We should also emphasize that the estimation is only based on model calculations. Although the QDCSM is proven successful in many situations, one can never claim the existence of the $N_{s\bar{s}}$ bound state unless it is discovered experimentally. Even if it exists, the mass and the width may also deviate from the model predictions. Together with the hidden charm pentaquark candidates discovered by LHCb, the investigations of this hidden strange pentaquark candidate $N_{s\bar{s}}$ may unravel the flavor-dependent properties and the structures of multiquark states. It will not only test the QDCSM and other phenomenological models, but also help advance our understanding of the strong interaction. In addition, the experimental search of the hidden strange pentaquark candidates is not a trivial extension of the hidden charm pentaquark states, because up to now almost all discovered multiquark states or candidates contain heavy quark components. The experimental exploration of light multiquark states is of unique significance to understand the structure of multiquark states. Therefore it deserves the efforts from both theories and experiments.

\acknowledgments{This work is supported in part by U.S. Department of Energy under Contract No. DEFG02-03ER41231 (H.G., T.L., Z.Z.). This work is also supported in part by the National Science Foundation of China under Contracts No. 11675080 (H.H.) and No. 11535005 (J.P., F.W., H.H.), the Natural Science Foundation of the  Jiangsu Higher Education Institutions of China under Grant No. 16KJB140006 (H.H.), and Jiangsu Government Scholarship for Overseas Studies (H.H.). It is also supported in part by the Duke Kunshan University (H.G., T.L.). We thank Dipangkar Dutta for providing the data of nucleon energy and momentum distributions in gold nucleus. }

\end{document}